\documentclass[aps,amssymb,amsmath,reprint]{revtex4-2}
\usepackage{graphicx,color}
\usepackage{epstopdf}
\usepackage{lineno}
\usepackage{comment}
\usepackage{soul}
\usepackage{ulem}
\usepackage{physics}
\usepackage{color}

\draft 

\begin{document}


\title{
Distinct memory properties in spin-wave reservoir computing based on synthetic antiferromagnet}


\author{Takumu Shinkai$^{1}$}
\affiliation{$^1$Department of Materials Physics, Nagoya University, Nagoya 464-8603, Japan}

\author{Satoshi Iihama$^{1}$}
\email{iihama.satoshi.y0@f.mail.nagoya-u.ac.jp}
\affiliation{$^1$Department of Materials Physics, Nagoya University, Nagoya 464-8603, Japan}

\author{Kensuke Hayashi$^{1}$}
\affiliation{$^1$Department of Materials Physics, Nagoya University, Nagoya 464-8603, Japan}

\author{Shigemi Mizukami$^{2,3}$}
\affiliation{$^2$WPI Advanced Institute for Materials Research (AIMR), Tohoku University, Sendai 980-8577, Japan}
\affiliation{$^3$Center for Science and Innovation in Spintronics (CSIS), Tohoku University, Sendai 980-8577, Japan}

\author{Natsuhiko Yoshinaga$^{4}$}
\affiliation{$^4$Department of Complex and Intelligent Systems, Future University Hakodate, Hokkaido 041-8655, Japan}

\author{Takahiro Moriyama$^{1}$}
\affiliation{$^1$Department of Materials Physics, Nagoya University, Nagoya 464-8603, Japan}



\date{\today}

\begin{abstract}
Spin-wave-based physical reservoir computing (RC) is a promising candidate for energy-efficient physical implementations of artificial intelligence because of its potential for nanoscale integration with low power consumption. 
Most of the previous studies on spin-wave RC have utilized spin waves excited in a single-layer ferromagnet.
In this study, we focused on spin waves in a synthetic antiferromagnet (SAF), consisting of two ferromagnetic layers coupled antiferromagnetically, and investigated additional memory properties of spin-wave RC.
We theoretically and numerically demonstrate the emergence of two distinct memory properties in the SAF device due to the distinct spin-wave characteristics of the acoustic and optical modes inherent in SAFs.

\end{abstract}
\maketitle

\section{Introduction}

Physical reservoir computing (RC) has attracted significant attention as a pathway towards energy-efficient physical implementations of artificial intelligence \cite{Tanaka2019, Nakajima2020}. 
Various physical systems have been proposed as physical RCs, using 
analog circuits\cite{Appeltant2011}, memristors\cite{Du2017, Moon2019, Milano2022, Liang2022, Chen2023}, photonics\cite{Vandoorne2014, Sande2017, Sugano2020, Rafayelyan2020, Nakajima2021}, and  spintronics\cite{Torrejon2017, Nakane2018, Prychynenko2018, Tsunegi2019, Grollier2020, Namiki2023, Tsunegi2023}.
Among various physical platforms, spintronics RC is a promising candidate because of its potential for nanoscale integration and low power consumption.
In particular, spin-wave-based physical RC has been intensively studied in recent years, in which spin waves are excited and detected by nanoscale spintronics technologies. 
Spin waves exhibit complex nonlinear dynamics and rich spatiotemporal characteristics, which are advantageous for physical RC.
In addition, spin waves can propagate over nanoscale distances at relatively low speeds, which is beneficial for achieving larger memory capacities at the nanoscale\cite{Iihama2024}. 
Most of the previous studies on spin-wave RC have utilized a single ferromagnetic layer\cite{Nakane2018, Nakane2023, Iihama2024, Furuta2018, Kanao2019, Watt2020, Watt2021, Namiki2023, Namiki2025, Nagase2024, Chen2025, Pinna2020, Lee2022}; however, such systems limit the degrees of freedom available to enhance memory capacities of RC.
When we use multilayers of magnetic materials such as synthetic antiferromagnets (SAFs), more complex spin-wave characteristics can be obtained due to the interaction between different magnetic layers.
SAF consists of two ferromagnetic layers coupled antiferromagnetically and supports two distinct eigenmodes, acoustic (AC) and optical (OP) modes, which exhibit different spin-wave characteristics and have rich physics\cite{Chiba2015, Kamimaki2020, Ishibashi2020, Sud2025}. 
If we utilize different spin waves with faster and slower modes in a single device, we can provide distinct memory properties with shorter and longer memory of input time-series, as schematically shown in the inset of Fig. 1(a), which might have potential to be used for prediction of time-series data with diverse timescale dynamics\cite{Zheng2020, Tanaka2022, Tokuda2024}.
In this study, we theoretically and numerically demonstrate the emergence of two distinct memory properties in SAF-based RC arising from the different spin-wave characteristics of the AC and OP modes inherent in SAFs.

\section{Setup of spin-wave RC using SAFs}

We performed micromagnetic simulations using the GPU-accelerated software MuMax$^3$\cite{Vansteenkiste2014}. 
Figure 1(a) shows a schematic illustration of spin-wave physical RC using the SAF, where spin waves are excited at the input node and spin-wave signals are detected at the output node.
Figure 1(b) shows the size and configuration of the system used in this study.
The model consists of a square SAF device with a system size of 1000 nm $\times $ 1000 nm, composed of two ferromagnetic layers coupled antiferromagnetically, with each ferromagnetic layer having a thickness of $d_{\rm F}$ = 5 nm.
The input and output nodes are symmetrically arranged along the center line with a separation distance $d$ = 800 nm as shown in Fig. 1(b), and the diameter of the node is set to $a$ = 40 nm.
Each ferromagnetic layer has a saturation magnetization of $M_{\rm s} = 1100$ kA/m, an exchange stiffness of $A_{\rm ex} = 1.1 \times 10^{-11}$ J/m, and a Gilbert damping constant of $\alpha = 0.01$.
The interlayer exchange coupling energy between the two ferromagnetic layers is set to $J_{\rm ex} = -1.7 \times 10^{-3}$ J/m$^2$.
These parameters assume a SAF consisting of the FeCoB / Ru / FeCoB structure\cite{Hashimoto2006, Cho2015, Watanabe2017, Kamimaki2020}. 
An external magnetic field ${\bf B}$ is applied along either the ${\bf x}$- or ${\bf y}$-directions.
The directions of the two magnetizations ${\bf m}_i$ are determined by the strength of ${\bf B}$ and the initial conditions chosen for ${\bf m}_i$.   
The z-component of the spin-polarized current proportional to the input time-series $U_n$ was injected into the input node to excite spin waves via the spin-transfer torque, {\it i. e.}, $j(t_n) = j_{\rm c} (U_n+1) /2$ where $U_n \in [-1, 1]$.
$j_{\rm c}$ was set to 5.0 $ \times 10^{11}$ A/m$^2$.
The damping constant $\alpha $ gradually increases up to 1 around the edge of the rectangle to prevent the reflection of spin waves [Fig. 1(b)]. 
The absorbing region was introduced outside a $850\,\mathrm{nm} \times 850\,\mathrm{nm}$ square. 
In this region, the damping constant increases quadratically in 10-nm-wide square shells, starting from $\alpha = 0.01$ at $850$--$860\,\mathrm{nm}$ and reaching $\alpha = 1.0$ at $950\,\mathrm{nm}$.
The z-component of magnetization was detected in the layer $i$, $m_{{\rm z}, i}$ at the output node, which is used to evaluate the memory properties of SAF-based RC.
The reservoir states ${\bf X}$ were constructed using the time-multiplexing technique\cite{Appeltant2011} as ${\bf X}(t_n) = \left[ m_{{\rm z}, i}(t_{n}+\theta), m_{{\rm z}, i}(t_{n}+2\theta), \cdots , m_{{\rm z}, i}(t_{n}+N_{\rm v}\theta) \right]$ where $\theta$ is the time interval for the time-multiplexing and $N_{\rm v}$ is the number of virtual nodes.
Three different combinations of output were used to construct ${\bf X}$, which are the magnetization of the first layer $m_{\rm z, 1}$, the summation of two layers $m_{\rm z, 1}+m_{\rm z, 2}$, and the difference $m_{\rm z, 1}-m_{\rm z, 2}$.  
Here, $\theta $ and $N_{\rm v}$ were set to 10 ps and 8, respectively.
The initial 200 time step calculations were discarded to eliminate the influence of the initial condition. 
The following 1800 time steps (= $N_{\rm T}$) were used for training and evaluation of the memory properties. 
The output weights ${\bf W}$ were trained to reconstruct the output time-series ${\bf Y}$ using linear regression as,
\begin{align}
  {\bf W} = {\bf Y} {\bf X}^{\dagger } .
\end{align}
To evaluate the memory property, delayed input time-series $U_{n-k}$ with delay $k$ was reconstructed from the spin-wave dynamics using linear regression\cite{Jaeger2002, Dambre2012}, {\it i. e.}, $Y_n = U_{n-k}$.
The capacity to reconstruct delayed input $C_k$ was calculated as
\begin{align}
C_k = \frac{\left< U_{n-k}, {\bf W}{\bf X}(t_n)\right>^2}{\left< U_{n-k}^2 \right>\left< \left( {\bf W}{\bf X}(t_n) \right)^2 \right>}.
\end{align}
Here, $\left< \cdots \right>$ represents the average over $n$.
$\left< a_n, b_n \right> $ is the covariance between $a_n$ and $b_n$.
The memory curve is obtained by plotting $C_k$ as a function of delay $k$.
The average delay $\left< k \right>$ of the memory curve can be calculated as
\begin{align}
   \left< k \right> = \frac{\sum_{k=1}^{k_{\rm max}} k C_k\Theta (C_k -\varepsilon )}{\sum _{k=1}^{k_{\rm max}}C_k \Theta (C_k -\varepsilon )},
\end{align}
where, $\varepsilon $ is the threshold to avoid overestimation of small $C_k$ values due to systematic positive errors and $\Theta (\cdot )$ is the Heaviside step function.
The value of $\varepsilon $ was set according to the method described in Ref. \cite{Dambre2012}.
Memory capacity ${\rm MC}$ is defined as
\begin{align}
    {\rm MC} = \sum_{k=1}^{k_{\rm max}} C_k \Theta (C_k -\varepsilon ).
\end{align}
Here, the maximum value that ${\rm MC} $ can take is $N_{\rm v}$\cite{Jaeger2002}.

\begin{figure}
\begin{center}
\includegraphics[width=0.45\textwidth,keepaspectratio,clip]{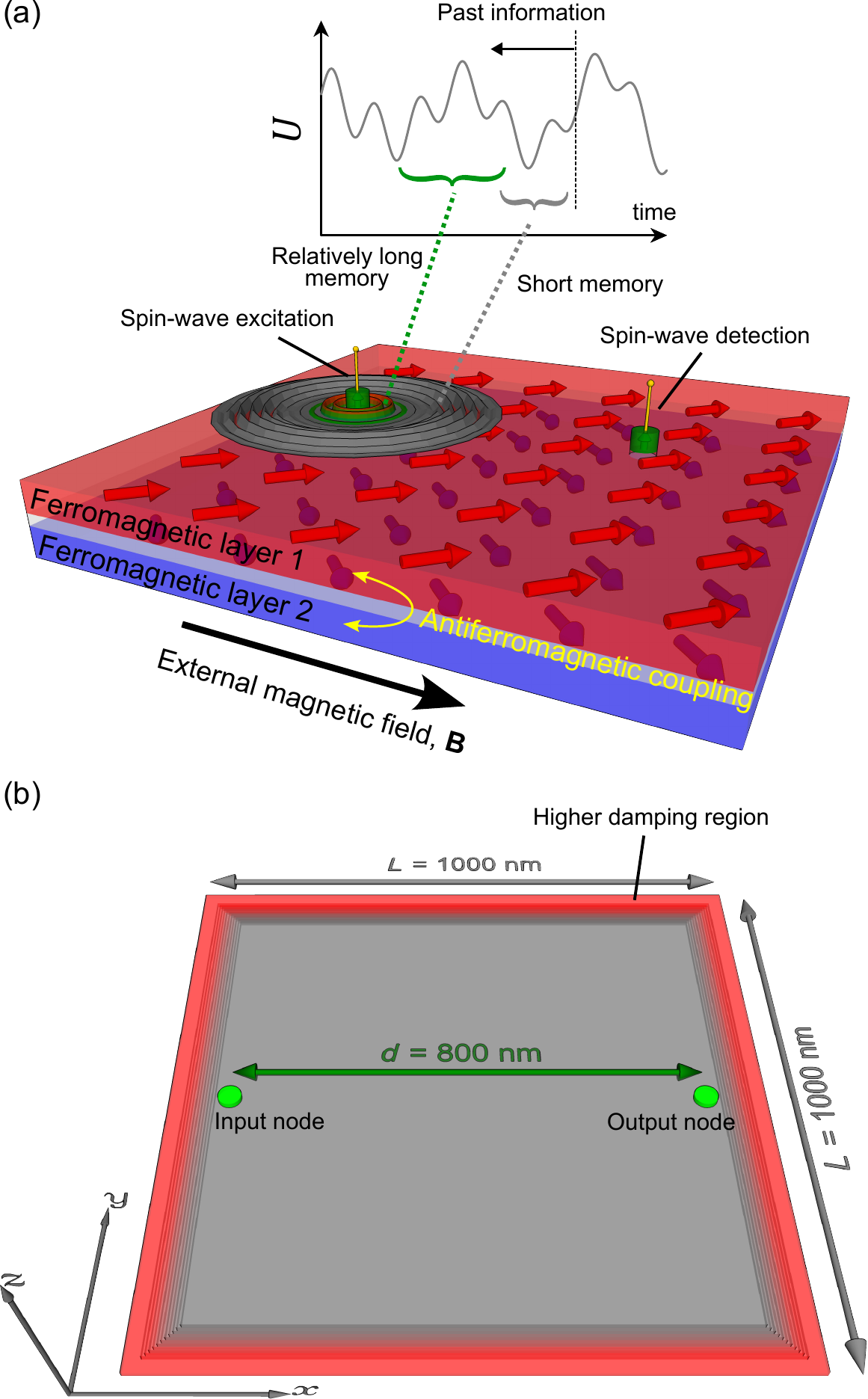}
\end{center}
\caption{(a)Schematic illustration of spin-wave physical reservoir computing using synthetic antiferromagnets. Magnetization dynamics is excited by spin-transfer torque to inject time-series data as spin waves and the spin-wave information is detected by magnetoresistance effect. Complex characteristics of spin-wave propagation in synthetic antiferromagnet enable to make distinct memory properties which can keep past input time-series with different time length. (b)Length scale to evaluate memory properties of spin-wave physical reservoir computing used in this study. Higher damping parameter is set at an edge of rectangle to prevent reflection of spin waves.}
\label{f1}
\end{figure}

\section{Results and discussion}

\begin{figure*}[t]
\begin{center}
\includegraphics[width=0.8\textwidth,keepaspectratio,clip]{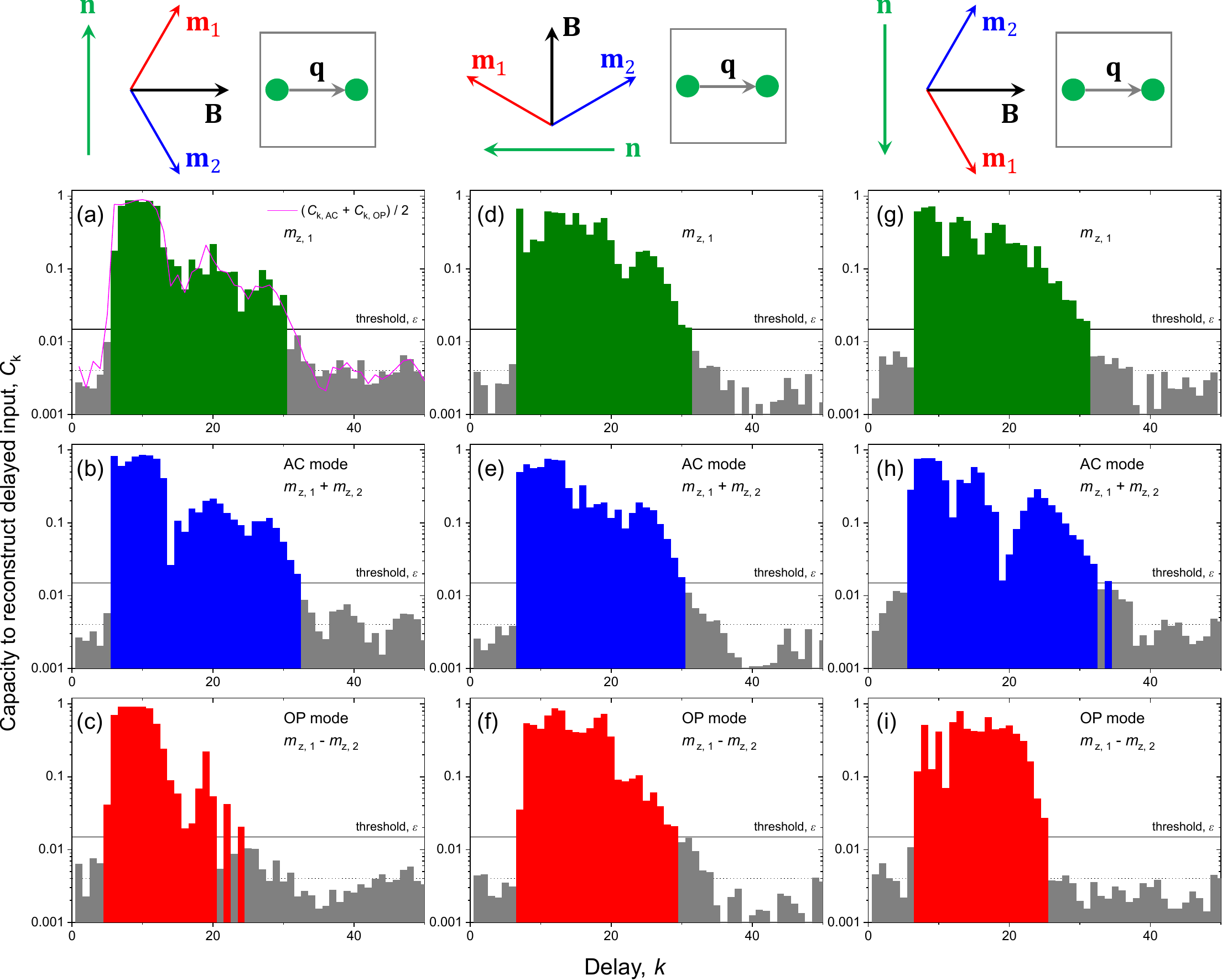}
\end{center}
\caption{ Memory curves of spin-wave physical reservoir computing using synthetic antiferromagnet. Capacity to reconstruct delayed input $C_k$ plotted as a function of delay $k$ of input time-series for the case when external magnetic field ${\bf B}$ and spin-wave wavevector ${\bf q}$ are parallel and N\'{e}el vector ${\bf n} = {\bf m}_1 -{\bf m}_2$ is along $+{\bf y}$ direction [(a)-(c)], the case when ${\bf B}$ and ${\bf q}$ are orthogonal [(d)-(f)], and the case when ${\bf B}$ and ${\bf q}$ are parallel and ${\bf n}$ is along $-{\bf y}$ [(g)-(i)]. Magnetization of one ferromagnetic layer ${\bf m}_1$ is used as output [(a), (d), (g)] while summations  ${\bf m}_1+{\bf m}_2$ [(b), (e), (h)] and differences ${\bf m}_1-{\bf m}_2$ [(c), (f), (i)] of two ferromagnetic layers are used. Here, $B$ = 0.2 T is applied. Solid curve in (a) is average value of $C_k$ obtained in (b) and (c). Solid and broken lines in (a)-(i) are the threshold $\varepsilon $ to avoid overestimation and $N_{\rm v}/N_{\rm T}$ expected value when there is no correlation between outputs and the data reconstructed by the magnetization. }
\label{f2}
\end{figure*}

\begin{figure}[t]
\begin{center}
\includegraphics[width=0.4\textwidth,keepaspectratio,clip]{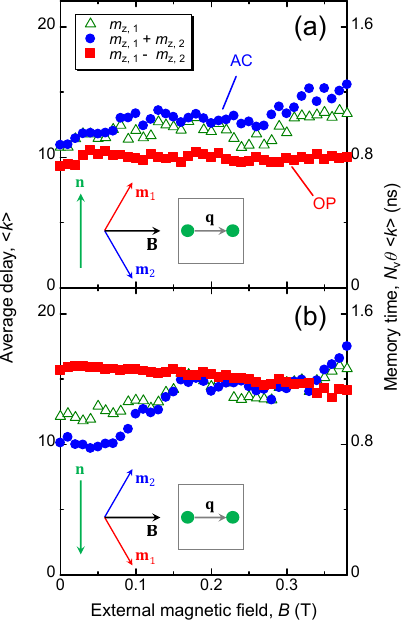}
\end{center}
\caption{(a) Average delay $\left< k \right>$ and memory time $N_{\rm v}\theta \left< k \right> $ plotted as a function of external magnetic field $B$ for the case when N\'{e}el vector ${\bf m}_1 -{\bf m}_2$ is along $+{\bf y}$ direction. (b) $\left< k \right>$ and $N_{\rm v}\theta \left< k \right> $ plotted as a function of $B$ when ${\bf m}_1 -{\bf m}_2$ is along $-{\bf y}$. }
\label{f3}
\end{figure}

Figure 2 shows the memory curves of SAF-based RC for three different configurations of the external magnetic field ${\bf B}$ and N\'{e}el vectors ${\bf n} = {\bf m}_1 -{\bf m}_2$ where ${\bf m}_1$ and ${\bf m}_2$ are magnetization of the two ferromagnetic layers.
Here, in-plane external magnetic field strength of $B$ = 0.2 T was applied.
When the delay $k$ exceeds approximately $k\sim $ 5, the values of $C_k$ exceed the threshold $\varepsilon $, which means that spin waves excited at the input node were propagated to the output node within the time $N_{\rm v}\theta k$.
After this delay $k\sim $ 5, the overall values of $C_k$ decrease with increasing $k$.
The shape of this memory curve depends on the configurations and combinations of the output.
The memory curve obtained with $m_{\rm z, 1}$ output can be reproduced by the average of memory curves for the in-phase AC mode ($m_{\rm z, 1}+m_{\rm z, 2}$) and the out-of-phase OP mode ($m_{\rm z, 1}-m_{\rm z, 2}$) [Solid curve in Fig. 2(a)].
The values of ${\rm MC}$ are found to be ${\rm MC} \sim N_{\rm v}$ in all configurations (See Appendix 1); however, the average delay values $\left< k \right>$ significantly differ depending on the configuration.
Figure 3 shows $\left< k \right>$ plotted as a function of the external magnetic field $B$ applied in the ${\bf x}$ direction with three different combinations of output and different N\'{e}el vector orientations.
$\left< k \right>$ with different configuration and output combinations for AC/OP modes revealed the presence of distinct memory properties in the single SAF-based RC device. 
When the N\'{e}el vector is along the $+{\bf y}$ direction, $\left< k \right>$ obtained with AC mode is higher than that evaluated with OP mode [Fig. 3(a)].
On the other hand, $\left< k \right>$ obtained in the AC mode is $\sim $1.5 times lower than that evaluated in the OP mode around $B$ $\sim $ 0.05 T when the N\'{e}el vector is along the $-{\bf y}$ direction [Fig. 3(b)].
These significant differences in $\left< k \right>$ depending on the N\'{e}el vector orientations and output combinations can be understood by considering the distinct spin-wave dispersion relations for AC and OP modes in SAFs.
When the spin-wave group velocity $v_{\rm g}$ is low, the time required for spin waves to propagate from the input to the output node increases.
Then, the spin waves can retain the input time-series for a longer time, resulting in larger $\left< k \right>$ and larger memory time $N_{\rm v}\theta \left< k \right>$.
In addition, we found that the difference in $\left< k \right>$ for AC and OP modes becomes less significant when the strength of the external magnetic field is increased (See Fig. 6 in Appendix 1).
This indicates that antiferromagnetic or non-collinear alignment of two ferromagnetic layers plays an important role for the emergence of distinct memory properties.

\begin{figure}
\begin{center}
\includegraphics[width=0.35\textwidth,keepaspectratio,clip]{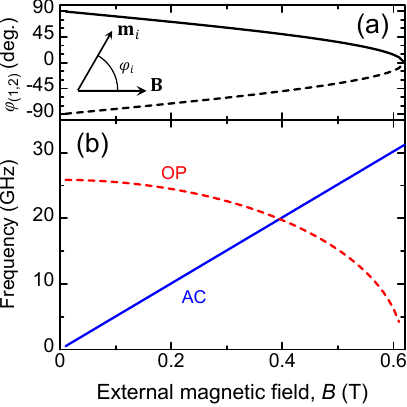}
\end{center}
\caption{(a)The relative angles $\varphi_{1}$ and $\varphi_{2}$ between $\mathbf{B}$ and $\mathbf{m}_{1,2}$. The solid and dashed lines represent $\varphi_{1}$ and $\varphi_{2}$, respectively. and (b) frequency at ${\bf q}$ = 0 for AC and OP modes calculated as a function of in-plane external magnetic field $B$.}
\label{f4}
\end{figure}

\begin{figure}
\begin{center}
\includegraphics[width=0.3\textwidth,keepaspectratio,clip]{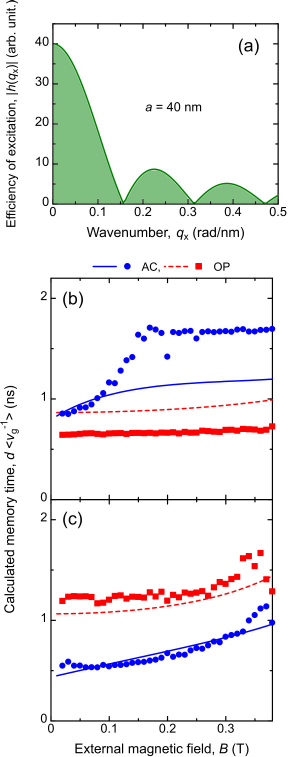}
\end{center}
\caption{(a) Efficiency of spin-wave excitation $|h(q_{\rm x})|$ plotted as a function of wavenumber $q_x$ with a parameter $a$ = 40 nm. (b), (c) Calculated memory time $d\left< v_{\rm g}^{-1}\right> $ for AC and OP modes based on the group velocity calculation of spin waves in SAFs with different N\'{e}el vector orientations (b) $+{\bf y}$ and (c) $-{\bf y}$ corresponding to Figs. 3(a) and 3(b). Solid and broken curves are the values evaluated with analytical calculation of spin-wave dispersion relation while circle and square symbols are the values evaluated with the dispersion relation obtained by micromagnetic simulations. }
\label{f5}
\end{figure}

To understand the difference in memory time for AC and OP modes, the spin-wave dispersion relation in SAFs and the average time required for spin waves to propagate from the input node to the output node $d\left< v_{\rm g}^{-1} \right>$ were calculated.
Figures 4(a) and 4(b) show the relative angle between ${\bf B}$ and ${\bf m}_{1, 2} $ and the spin-wave frequency at ${\bf q}$ = 0 for AC and OP modes calculated as a function of the in-plane external magnetic field $B$.
As $B$ increases, $\varphi _{1, 2} $ approaches to 0 deg. and two ferromagnetic layers aligned parallel eventually at around $B$ = 0.6 T. 
The spin-wave dispersion relation in SAFs was obtained by analytical calculation considering the magneto-static spin wave, the exchange spin wave, and the dynamic dipolar coupling between two ferromagnetic layers and by calculating a Fourier transformation of the data using micromagnetic simulation (See Appendix 3 for calculation details).
Spin waves are more efficiently excited near the spin-wave wavevector $q_{\rm x} \sim $ 0 due to the limited size of the input node.
Thus, $\left< v_{\rm g}^{-1} \right>$ can be calculated as
\begin{align}
    \left< v_{\rm g}^{-1} \right> = \frac{1}{A} \int ^{\infty }_0 v_{\rm g}^{-1}(q_{\rm x}) \left| h (q_{\rm x}) \right| {\rm d}q_{\rm x} , \label{eq:vg-1}
\end{align}
where, $h(q_{\rm x})$ is the spin-wave excitation efficiency determined by the spatial profile of the input node and $A$ is the normalization factor, which are respectively given by
\begin{align}
   &h(q_{\rm x}) = \frac{2h_0}{q_{\rm x}}\sin \left( \frac{q_{\rm x} a}{2}\right) , \label{eq:hq} \\
  &A = \int ^{\infty }_0 \left| h(q_{\rm x}) \right| {\rm d}q_{\rm x} . \label{eq:A} 
\end{align}  
The shape of $|h (q_{\rm x})|$ is shown in Fig. 5(a).
$v_{\rm g} $ is the slope of the dispersion relation, {\it i. e.}, $v_g$ = ${\rm d}(2\pi f)/{\rm d}q_{\rm x}$. 
The dispersion relation of spin waves in SAFs was obtained using the parameters used in the micromagnetic simulations.
The slope of the dispersion relation for AC and OP modes around $q_{\rm x} \sim 0$ differs due to the non-reciprocal spin waves caused by the dynamic dipolar coupling between two ferromagnetic layers\cite{Grunberg1985, Kabos1994, Di2015, Gallardo2019, Ishibashi2020, Millo2023}.
The ordering of $v_{\rm g}$ at $q_{\rm x} \sim 0$ between AC and OP modes was reversed by the N\'{e}el vector orientations as a result of spin-wave non-reciprocity (See Figs. 9(c) and 9(d) in Appendix 3).
Figures 5(b) and 5(c) show the memory time $d\left< v_{\rm g}^{-1} \right>$ calculated using Eq. (\ref{eq:vg-1}) plotted as a function of external magnetic field $B$ for different N\'{e}el vector orientations.
The values of $v_{\rm g}$ larger than a critical value of $v_{\rm g, c}$ = 100 m/s are used to calculate the integral in Eq. (\ref{eq:vg-1}) to avoid divergence of $v_{\rm g}^{-1}$ and reduce numerical errors.
The calculated memory time $d \left< v_{\rm g}^{-1} \right> $ reproduces the trend of the memory time $N_{\rm v}\theta \left< k \right>$ obtained for spin-wave RC [Fig. 3], indicating that the distinct memory properties in SAF-based RC originate from the different spin-wave characteristics of the AC and OP modes.
The slight deviations in $d\left< v_{\rm g}^{-1}\right> $ in Figs. 5(b) and 5(c) are caused by the differences between the analytical and numerical evaluation of the dispersion relations (See details in Appendix 3).

Although MC has been widely studied, the distribution of memory curves has been focused recently\cite{Iacob2024}.
However, the indicators of the memory curve, such as $\left< k \right>$, have not been reported in spin-wave RC.
The consideration of $\left< k \right>$ in the memory property for spin-wave RC described above can also be used for the case of single-layer ferromagnetic thin films (See Appendix 2). 
Designing the memory properties considering spin-wave characteristics and input-output configurations will provide a pathway to enhance the performance of spin-wave RC for various time-series prediction tasks.

This study presents only the case of linear memory properties; however, distinct memory properties in SAF-based spin-wave RC might also provide information processing capacity, {\it i. e.}, nonlinear memory properties, if we use nonlinear magnetization dynamics in SAFs\cite{Kamimaki2020, He2025, Sud2025, Mouhoub2026, You2026}.
Understanding and utilizing nonlinear magnetization dynamics for spin-wave RC has remained an issue, which will be subjects for future works.

\section{Conclusion}

In this study, we demonstrated the emergence of two distinct memory properties in SAF-based RC arising from the different spin-wave characteristics of the AC and OP modes inherent in SAFs.
$\left< k \right>$ in the memory curve for spin-wave RC were evaluated using micromagnetic simulations, and significant differences were found depending on the N\'{e}el vector orientations and output combinations.
The differences in memory time were explained by the different group velocity of spin waves for AC and OP modes, which is determined by the spin-wave dispersion relation in SAFs.
The present results indicate that multilayered magnetic systems provide a pathway to enhance the degrees of freedom of spin-wave RC and to achieve distinct memory properties in a single device, which may have the potential to be used for more complex time-series prediction including diverse timescale dynamics.

\begin{acknowledgments}
This work was supported by JST PRESTO (No. JPMJPR22B2), JST FOREST (No. JPMJFR2242, No. JPMJFR2140), JST CREST (No. JPMJCR24R5) and KAKENHI (No. 24K21234, No. 24H02235, No. 24K23945). S. M. acknowledges Spin-RNJ.
\end{acknowledgments}

\section*{Data Availability}
The data that support the findings of this study are available from the corresponding author upon reasonable request.

\appendix

\section*{Appendix}

\subsection{Memory capacities and average delays for various configurations}

Figure 6 shows memory capacities MC and average delays $\left< k \right>$ for AC and OP modes plotted as a function of external magnetic field with three different magnetization configurations.
MC for all configurations and both AC and OP modes are found to be $\sim N_{\rm v}$, indicating that the decay of the spin-wave amplitude during propagation does not significantly affect the memory property in the current system size and time steps used for RC.
On the other hand, $\left< k \right>$ values differ significantly depending on the configuration especially at low magnetic field.
This is attributed to the different spin-wave characteristics of AC and OP modes in SAFs as discussed in the main text.

\begin{figure}[h]
\begin{center}
\includegraphics[width=0.48\textwidth,keepaspectratio,clip]{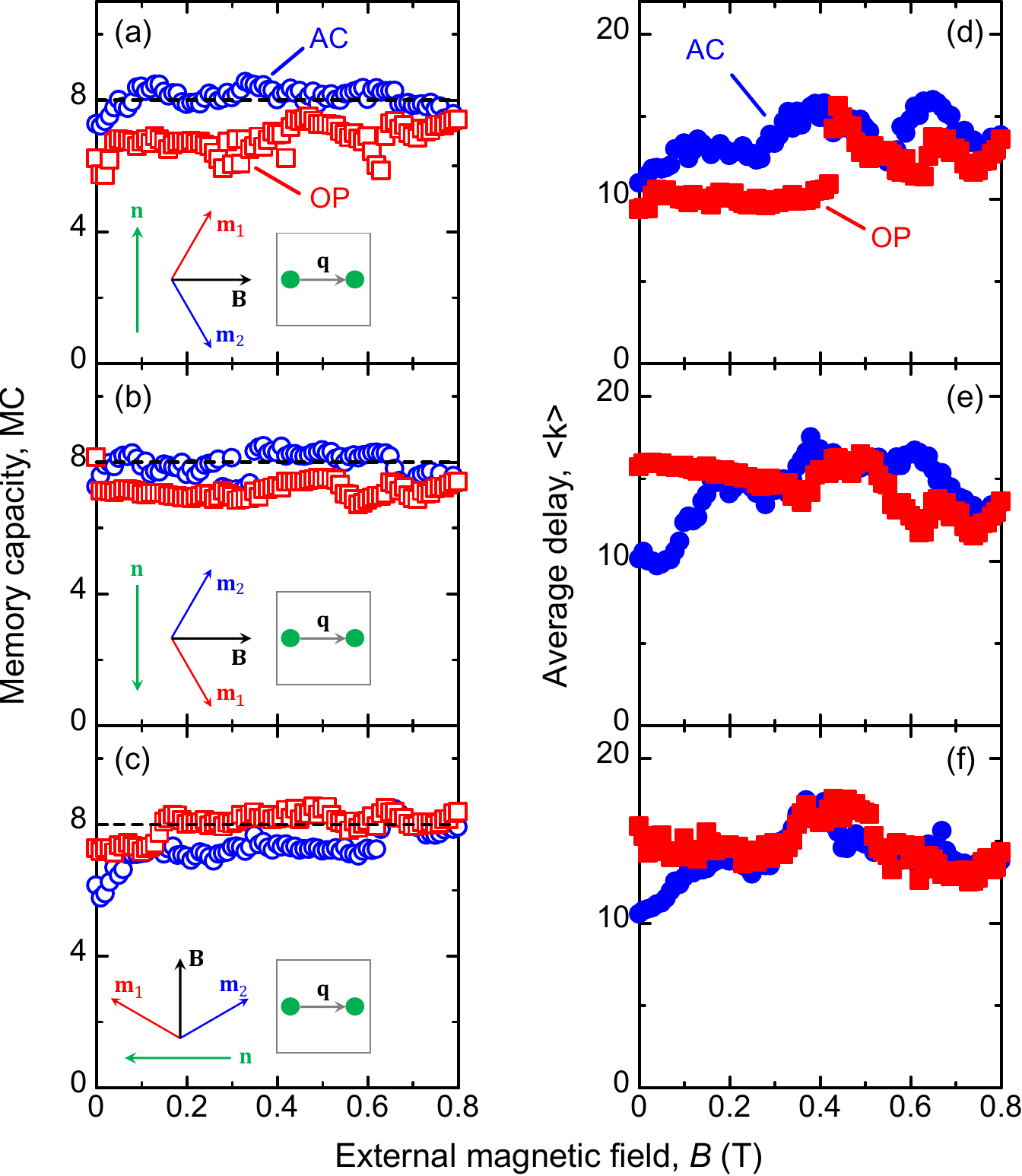}
\end{center}
\caption{(a), (b), (c) MC for AC and OP modes plotted as a function of external magnetic field with three different magnetization configurations. Broken line indicates the number of virtual nodes $N_{\rm v}$. (d), (e), (f) $\left< k \right>$ for AC and OP modes plotted as a function of external magnetic field $B$ with three different magnetization configurations. }
\label{fA1}
\end{figure}

\subsection{Case for the single layer ferromagnetic thin film}

\begin{figure}
\begin{center}
\includegraphics[width=0.35\textwidth,keepaspectratio,clip]{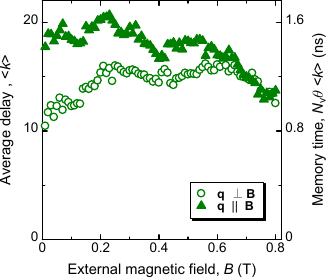}
\end{center}
\caption{Average delay $\left< k\right> $ for the ferromagnetic thin film  plotted as a function of external magnetic field $B$ with different configurations. }
\label{f6}
\end{figure}

\begin{figure}
\begin{center}
\includegraphics[width=0.3\textwidth,keepaspectratio,clip]{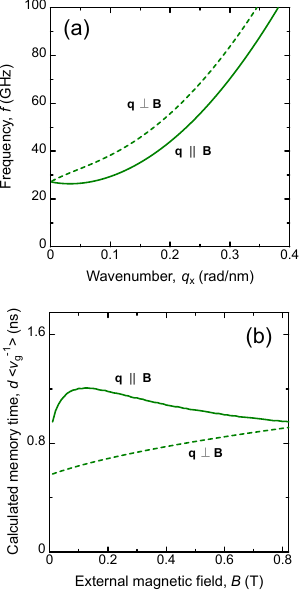}
\end{center}
\caption{(a) Calculated spin-wave dispersion relation of ferromagnetic thin film with different configuration. Strength of external magnetic field $B$ used in the calculation is 0.5 T. (b) Calculated memory time $d \left< v_{\rm g}^{-1}\right> $ plotted as a function of external magnetic field $B$ with two different configurations. }
\label{f7}
\end{figure}

\begin{figure*}
\begin{center}
\includegraphics[width=0.7\textwidth,keepaspectratio,clip]{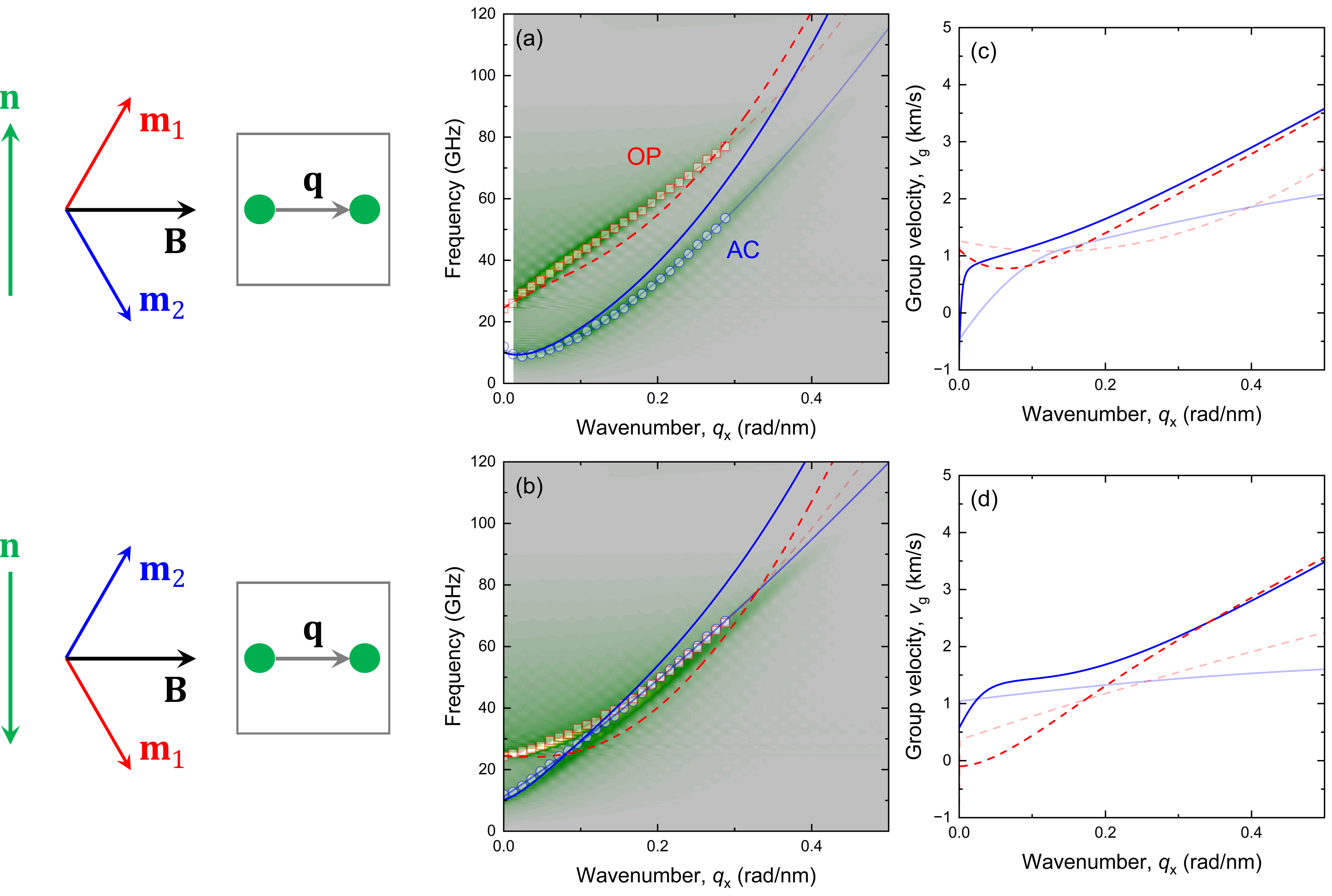}
\end{center}
\caption{(a), (b) Spin-wave frequency $\omega /(2\pi )$ for acoustic (AC) and optical (OP) mode of synthetic antiferromagnet plotted as a function of $q_{\rm x}$ when N\'{e}el vector ${\bf m}_1 -{\bf m}_2$ is along (a) $+{\bf y}$ and (b) $-{\bf y}$. Here, external magnetic field is applied along ${\bf x}$ direction with the strength of $B$ = 0.2 T. Solid and broken curves are the dispersion relations obtained by the analytical calculation while color maps are the dispersion relation obtained by micromagnetic simulations. Peak values of the color map at each wavenuber $q_{\rm x}$ are extracted as shown in circle and square symbols, which are complemented by spline smoothing as shown in thin solid and broken curves. (c), (d) Group velocity of spin-wave $v_{\rm g}$ obtained by the slope of dispersion relations shown in Figs. (a) and (b). }
\label{f7}
\end{figure*}

The average delay $\left< k\right> $ and the corresponding memory time $N_{\rm v}\theta \left< k \right>$ for single layer ferromagnetic thin film were also evaluated with the same procedure as described in the main text. 
The thickness of the ferromagnetic layer was set to $d_{\rm F}$ = 10 nm, and other parameters were fixed the same as in the case for the SAF.
Figure 7 shows $\left< k \right>$ plotted as a function of $B$ with two different configurations, ${\bf q} \perp {\bf B}$ and ${\bf q} \parallel {\bf B}$.
$\left< k \right>$ for ${\bf q} \perp {\bf B}$ configuration is smaller than that for ${\bf q} \parallel {\bf B}$ configuration, which is possibly caused by different $v_{\rm g}$ of two configurations in spin-wave mode typically called Damon-Eshbach (DE) mode and backward volume (BV) mode.
The dispersion relation for the DE and BV modes can be expressed as\cite{Hillebrands2007},
\begin{align}
  & \omega _{\rm DE} = \sqrt{\left( \omega _{\rm B} + \omega _{\rm ex} \right)\left( \omega _{\rm B} + \omega _{\rm ex} + \omega _{\rm M}  \right)+\frac{\omega_{\rm M}^2}{4} \left( 1- e^{-2qd_{\rm F}}\right)} , \\
  & \omega _{\rm BV} = \sqrt{\left( \omega _{\rm B} + \omega _{\rm ex} \right)\left( \omega _{\rm B} + \omega _{\rm ex} + \omega _{\rm M} N_{\rm q}  \right)} ,
\end{align}
where, $\omega _{\rm B}$, $\omega _{\rm M}$, and $\omega _{\rm ex}$ are angular frequency due to the external magnetic field, demagnetizing field, and exchange stiffness in the ferromagnetic layer, which are respectively given by $ \omega _{\rm B} = \gamma B$ and $\omega _{\rm M} = \gamma \mu _0 M_{\rm s}$, and $\omega _{\rm ex} = \frac{2A_{\rm ex}}{M_{\rm s}} \gamma  q^2$.
$\gamma $ is the gyromagnetic ratio.
$N_{\rm q}$ represents the wavenumber dependent demagnetizing factor, which is given by $N_{\rm q} = (1-e^{-qd_{\rm F}})/(qd_{\rm F})$.
Figure 8(a) shows the calculated dispersion relation ($\omega _{\rm DE}/(2\pi )$ and $\omega _{\rm BV}/(2\pi )$) with $B$ = 0.5 T.
The slope of the dispersion relation for the BV mode is smaller than that for the DE mode around $q _{\rm x} \sim 0$, which is attributed to the competition between the negative slope of the BV mode and the exchange spin wave.
As a result, the memory time $d\left< v_{\rm g}^{-1} \right>$ for the BV mode is larger than that for the DE mode, as shown in Fig. 8(b), which is consistent with the $\left< k \right>$ values obtained for two different configurations in spin-wave modes for the single-layer ferromagnetic thin film [Fig. 7].

\subsection{Spin-wave dispersion relation of SAF}

The dispersion relation of spin waves in SAFs was obtained considering eigenfrequency analysis of the Landau-Lifshitz-Gilbert (LLG) equation of coupled two ferromagnetic layers, similar to the method described in \cite{Shiota2020}.
The LLG equation without considering the damping term is given by
\begin{align}
    \frac{{\rm d}{\bf m}_i}{{\rm d}t} = -\gamma {\bf m}_i \times {\bf B}_{{\rm eff}, i} \label{eq:LLG}
\end{align}
where ${\bf B}_{{\rm eff}, i}$ is the effective magnetic field for the layer $i$.
$i, j \in \{1,2\}$, $i\neq j$ are the indexes for two ferromagnetic layers.
The external magnetic field, the interlayer exchange coupling, the exchange stiffness in the layer, and the magnetic dipolar field are considered in ${\bf B}_{{\rm eff}, i}$ in Eq. (\ref{eq:LLG}).
Eq. (\ref{eq:LLG}) was linearized around the equilibrium configuration by considering the terms linear in ${\bf m}_i$.
After performing a Fourier transformation with respect to time, {\it i. e.} $\frac{{\rm d}{\bf m}_i}{{\rm d}t} \rightarrow {\rm i} \omega {\bf m}_i$, one can obtain the following equation in matrix form as
\begin{align}
  \left( 
  \begin{array}{cccc}
    \omega _{{\rm X}_1{\rm X}_1} & -{\rm i} \omega & \lambda _{{\rm X}_1{\rm X}_2} & {\rm i} \zeta \\
    {\rm i} \omega & \omega _{{\rm Y}_1{\rm Y}_1} & {\rm i} \eta  & \lambda _{{\rm Y}_1{\rm Y}_2} \\
    \lambda _{{\rm X}_2{\rm X}_1} & -{\rm i} \eta & \omega _{{\rm X}_2{\rm X}_2} & -{\rm i} \omega \\
    -{\rm i} \zeta & \lambda _{{\rm Y}_2{\rm Y}_1} & {\rm i} \omega & \omega _{{\rm Y}_2{\rm Y}_2}
  \end{array}
  \right)
  \left(
  \begin{array}{c}
    m_{{\rm X}_1} \\
    m_{{\rm Y}_1} \\
    m_{{\rm X}_2} \\
    m_{{\rm Y}_2}
  \end{array}
  \right) = {\bf 0} . \label{eq:matrix}
\end{align}
Here, $\omega _{{\rm X}_i{\rm X}_i}$, $\omega _{{\rm Y}_i{\rm Y}_i}$, $\lambda _{{\rm X}_i{\rm X}_j}$, $\lambda _{{\rm Y}_i{\rm Y}_j}$, $\eta $ and $\zeta $ are given by,
\begin{widetext}
\begin{align}
  &\omega _{{\rm X}_i{\rm X}_i} = \omega _{\rm B} \cos (\varphi _i - \varphi _{\rm B}) + \omega _{\rm E} \cos (\varphi _i -\varphi _j) +\omega _{\rm ex} + \omega _{\rm M} (1 - N_{\rm q}) \left( \cos ^2 \varphi _{\rm q} \sin ^2 \varphi _i - \frac{1}{4} \sin 2 \varphi _{\rm q} \sin 2 \varphi _i \right) , \notag \\
  &\omega _{{\rm Y}_i{\rm Y}_i} = \omega _{\rm B} \cos (\varphi _i - \varphi _{\rm B}) + \omega _{\rm E} \cos (\varphi _i -\varphi _j) + \omega _{\rm ex} + \omega _{\rm M} N_{\rm q} , \notag \\  
  &\lambda _{{\rm X}_i{\rm X}_j} = \lambda _{{\rm X}_j{\rm X}_i} = -\omega _{\rm E} \cos (\varphi _i -\varphi _j) + \omega _{\rm M} \xi _{\rm q} \left( \cos ^2 \varphi _{\rm q} \sin \varphi _i \sin \varphi _j + \sin ^2 \varphi _{\rm q} \cos \varphi _i \cos \varphi _j  - \frac{1}{2} \sin 2 \varphi _{\rm q} \sin (\varphi _i + \varphi _j ) \right) , \notag \\
  &\lambda _{{\rm Y}_i{\rm Y}_j} = \lambda _{{\rm Y}_j{\rm Y}_i} = -\omega _{\rm E} + \omega _{\rm M} \xi _{\rm q} , \quad \eta = \omega _{\rm M} \xi _{\rm q} \sin (\varphi _1 - \varphi _{\rm q}) , \quad  \zeta = \omega _{\rm M} \xi _{\rm q} \sin (\varphi _2 - \varphi _{\rm q}) , \notag
\end{align}
\end{widetext}
where, $\omega _{\rm E}$ is the angular frequency due to the exchange coupling between layers, which is given by $\omega _{\rm E} = \gamma J_{\rm ex}/(M_{\rm s} d_{\rm F})$.
${\rm X}_i$, ${\rm Y}_i$ represent two orthogonal components of the coordinate with respect to ${\bf m}_i$.
$\varphi _{\rm B}$ and $\varphi _{\rm q}$ are in-plane angles for the external magnetic field and the spin-wave wavevector, respectively.
The condition $\varphi _{\rm B} = (\varphi _1 + \varphi _2) /2 $ is satisfied due to the structural symmetry of the SAF.
$\xi _{\rm q}$ represents the wavenumber dependent strength of the dynamic dipolar coupling between two layers, which is given by, $\xi _{\rm q}= 2e^{-qd_{\rm F}}\sinh ^2(qd_{\rm F}/2)/(qd_{\rm F})$.
The determinant of the 4$\times $4 matrix shown in Eq. (\ref{eq:matrix}) gives the resonance condition depending on the wavenumber $q$ and its angle $\varphi _{\rm q}$, which leads to the following quartic equation in $\omega $ as
\begin{align}
\omega ^4 + c_2 \omega ^2 + c_1 \omega + c_0 = 0 . \label{eq:omega}
\end{align}
This quartic equation gives rise to different positive and negative solutions in $|\omega |$, which is a signature of nonreciprocity.
When the condition that N\'{e}el vector is orthogonal to ${\bf q}$, $\eta $ becomes equal to $\zeta $.
This condition leads to $c_1$ = 0, which indicates that $|\omega |$ in positive and negative solutions is the same, resulting in a reciprocal dispersion relation.

In addition to the analytical calculation as mentioned above the dispersion relation of spin waves in SAFs was also obtained by performing a Fourier transformation of the magnetization calculated using micromagnetic simulations.
The Fourier transformation spectra were then analyzed to extract the peak frequency at each wavenumber $q_{\rm x}$, which was complemented by spline smoothing to obtain $v_{\rm g}$.

Figures 9(a) and 9(b) show the dispersion relation of spin waves in SAFs with different N\'{e}el vector orientations.
The dispersion relation obtained by the analytical calculation more or less coincides with that obtained by micromagnetic simulations. 
The dispersion relations between two different N\'{e}el vector polarities especially at $q_{\rm x} $ $\sim $ 0 differ due to the nonreciprocal spin waves.
Figures 9(c) and 9(d) show corresponding $v_{\rm g}$, the slope of the dispersion relations shown in Figs. 9(a) and 9(b).
The spin waves at $q_{\rm x} $ $\sim $ 0 are more efficiently excited [Fig. 5(a)] where the $v_{\rm g}$ difference between the AC and OP modes is large, which gives rise to distinct memory properties in SAFs.
The deviations in $d\left< v_{\rm g}^{-1}\right> $ in Figs. 5(b) and 5(c) are caused by the differences between the analytical and numerical evaluation of the dispersion relations and its slopes shown in Figs. 9(a, b) and 9(c, d).  

\bibliography{export}

\end{document}